\documentstyle [twocolumn,epsf]{mn}
\newcommand{\mincir}{\raise
-2.truept\hbox{\rlap{\hbox{$\sim$}}\raise5.truept\hbox{$<$}\ }}
\newcommand{\magcir}{\raise
-2.truept\hbox{\rlap{\hbox{$\sim$}}\raise5.truept\hbox{$>$}\ }}
\newcommand{\minmag}{\raise
-2.truept\hbox{\rlap{\hbox{$<$}}\raise6.truept\hbox{$<$}\ }}
\newcommand{\be}{\begin{equation}}
\newcommand{\ee}{\end{equation}}
\newcommand{\ba}{\begin{eqnarray}}
\newcommand{\ea}{\end{esqnarray}}
\newcommand{\brr}{\begin{array}}
\newcommand{\err}{\end{array}}
\newcommand{\bc}{\begin{center}}
\newcommand{\ec}{\end{center}}

\renewcommand{\baselinestretch}{1.2}
\title[Virialization of cosmological structures in models with time 
varying equation of state]{Virialization of cosmological structures 
in models with time varying equation of state}
\author[Spyros Basilakos \& Nikos Voglis]
{Spyros Basilakos \& Nikos Voglis \\
Academy of Athens, Research Center for Astronomy \& Applied
  Mathematics, Soranou Efessiou 4, 11-527, Athens, Greece\\
}

\begin{document}

\maketitle

\begin{abstract}
We study the virialization 
of the cosmic structures 
in the framework of flat cosmological models where the dark energy 
component plays an important role in the 
global dynamics of the universe.
In particular, our analysis focuses on the study of the spherical 
matter perturbations,
as the latter decouple from the background expansion and start
to ``turn around'' and finally collapse. We generalize this 
procedure taking into account 
models with an equation of state which vary with time, and
provide a complete formulation of the
cluster virialization 
attempting to address the
nonlinear regime of structure formation. In particular,
assuming that clusters have collapsed prior to the epoch 
of $z_{\rm f}\simeq 1.4$, in which 
the most distant cluster has been found, 
we show that the behavior of the spherical collapse model 
depends on the functional form of the equation of state.

\vspace{0.25cm}

\noindent
{\bf Keywords}: 
clusters: formation-
cosmology:theory - large-scale structure of universe
\end{abstract}

\section{Introduction}
It is well known that the available high quality 
cosmological data (Type Ia supernovae, 
CMB, etc.) are well fitted by an emerging ``standard model'', which
contains cold dark matter (CDM) to explain clustering and an extra component
with negative pressure, the dark energy, 
to explain the observed accelerated cosmic expansion.
The last decades there have been many theoretical 
speculations regarding the nature 
of the above exotic dark energy. Most of the 
authors claim that 
a scalar field which rolls down the potential $V(\phi)$ 
(Ratra \& Peebles 1988; Caldwell, Dave, Steinhardt 1998; 
Peebles \& Ratra 2003) could resembles 
that of the dark energy.
The cosmic equation of state is given by $p_{Q}=w\rho_{Q}$,  
where $p_{Q}$ is the pressure and $\rho_{Q}$ is the corresponding
density of the dark energy. Owing to the fact that 
$w<-1/3$ is related to the potential of the
dark energy field, having an indication about its value  
may help us to understand the nature of the dark energy.

However, a serious issue here is how the large scale
structures and in particular galaxy clusters form. The 
cluster distribution traces basically scales that have 
not yet undergone the non-linear phase of gravitationally 
clustering, thus simplifying their connections
to the initial conditions of cosmic structure formation. 
The so called spherical collapse model, which has a long history
in cosmology, is a simple but still a fundamental tool to
describe the growth of bound systems in the universe via 
gravitation instability
(Gunn \& Gott 1972). In the last decade
many authors have been involved in this kind of studies
and have found that the main parameters 
of the spherical collapse model
such as the 
ratio between the final and the turn around radius (hereafter collapse
factor), 
is affected by the dark energy
(Lahav et al. 1991; Wang \& Steinhardt 1998; Iliev \& Shapiro 2001;
Battye \& Weller 2003; Basilakos 2003; Weinberg \& Kamionkowski 2003;
Mota \& van de Bruck 2004; Horellou \& Berge 2005; 
Zeng \& Gao 2005; Maor \& Lahav 2005; Percival 2005; Wang 2006; Nunes
\& Mota 2006). 

The aim of this work is along the same lines, attempting 
to investigate the cluster formation processes
by generalizing the non-linear spherical model for a family of 
cosmological models with 
an equation of state parameter being a function of time, $w=w(t)$.
This can help us  
to understand better the theoretical expectations of
negative pressure models as well as the variants from the Quintessence
($w=w_{0}=const$ with $-1\le w_{0}<-1/3$) and Phantom
($w=w_{0}=const$ with $w_{0}<-1$) 
case respectively.

The structure of the paper is as follows. 
The basic elements of the cosmological equations are presented in 
section 2. Sections 3
and 4 outline the spherical collapse analysis 
in models where $w$ is a function of time and
finally, we draw our conclusions in section 5.

\section{The basic Cosmological Equations}
For homogeneous and isotropic cosmologies, driven by 
non relativistic matter and an
exotic fluid with equation of state, 
$p_{Q}=w(\alpha)\rho_{Q}$
with $p_{Q} <0$, the Einstein field equations can be 
given by:
\begin{equation}
\left( \frac{\dot{\alpha}}{\alpha} \right)^{2}=
\frac{8\pi G}{3}(\rho_{\rm m}+\rho_{Q})-\frac{k}{\alpha^{2}}
\end{equation}
with $k=-1, 0$ or 1 for
open, flat and closed universe respectively
and 

\begin{equation}
\frac{\ddot{\alpha}}{\alpha}=-4\pi G
\left[\left(w(\alpha)+\frac{1}{3}\right)\rho_{Q}+\frac{1}{3}\rho_{\rm m}\right] \;\;,
\end{equation}
where $\alpha(t)$ is the scale 
factor, $\rho_{\rm m}=\rho_{\rm m0} \alpha^{-3}$ is the background
matter density and 
$\rho_{Q}=\rho_{Q0} f(\alpha)$ is the dark 
energy density, with:
\begin{equation}
f(\alpha)={\rm exp}\left[ 3\int_{\alpha}^{1} \left(
  \frac{1+w(u)}{u}\right) {\rm d}u \right] \;\;.
\end{equation}

Thus, the scale 
factor evolves according to Friedmann equation:
$H^{2}\equiv (\dot{\alpha}/\alpha)^{2}$.
The relation between the time and the scale factor 
is given by 
\begin{equation}
\frac{dt}{d\alpha}=\frac{1}{H(\alpha) \alpha} \;\;,
\end{equation}
where the Hubble parameter is written: 
$H(\alpha)=H_{0}E(\alpha)$ 
and $H_{0}$ is the Hubble constant with
\begin{equation}
E(\alpha)=\left[ \Omega_{\rm m}\alpha^{-3}+\Omega_{k}\alpha^{-2}+
\Omega_{Q}f(\alpha)\right]^{1/2} \;\;\;,
\end{equation}
while $\Omega_{\rm m}= 8\pi G \rho_{\rm m0}/3H_{0}^{2}$ 
(density parameter), $\Omega_{k}=-k/H_{0}$ (curvature parameter),
$\Omega_{Q}= 8\pi G \rho_{Q0}/3H_{0}^{2}$ 
(dark energy parameter) at the present time with
$\Omega_{\rm m}+\Omega_{k}+\Omega_{Q}=1$.

In addition, to $\Omega_{m}(\alpha)$ also $\Omega_{Q}(\alpha)$ could evolve 
with the scale factor as
\begin{equation}
\Omega_{\rm m}(\alpha)=\frac{\Omega_{\rm m} 
\alpha^{-3}}{E^{2}(\alpha)} \;\;\;{\rm and} \;\;\; 
\Omega_{Q}(\alpha)=\frac{\Omega_{Q} f(\alpha)}{E^{2}(\alpha)} \;\; .
\end{equation}
Note that in this paper we consider a spatially flat ($k=0$) 
low-$\Omega_{\rm m}=0.3$ cosmology with
$H_{0}=72$ Km s$^{-1}$Mpc$^{-1}$ which is in agreement with 
the cosmological parameters found from the recent observations 
(see Freedman et al. 2001; Spergel et al. 2003; Tegmark et al. 2004; 
Basilakos \& Plionis 2005; Spergel et al. 2006 and references therein).
Finally, in order to address the negative pressure term it is
essential to define the functional form of the 
equation parameter $w=w(\alpha)$ [see section 4].

\section{Virialization in the Spherical Model}
The spherical collapse model is still a powerful tool, despite its
simplicity, for understanding how a small spherical patch of
homogeneous overdensity forms a bound system via 
gravitation instability. Technically speaking,
the basic cosmological equations, mentioned before, are valid either for 
the entire universe or for homogeneous spherical perturbations 
[by replacing the scale factor with radius $R(t)$] 
\begin{equation}
\frac{\ddot{R}}{R}=-4\pi G\left[\left(w(R)+\frac{1}{3}\right)\rho_{\rm
    Qc}+
\frac{1}{3}\rho_{\rm mc}\right] \;\;,
\end{equation}
where $\rho_{\rm mc}$ and $\rho_{\rm Qc}$ is the time-varying 
matter and dark energy density respectively (for spherical 
perturbations).

We study the cluster virialization in models with dark energy,
generalizing the notations of 
Lahav et al. (1991), Wang \& Steinhardt (1998), Mainini et al. (2003);
Basilakos (2003), Lokas (2003) 
Mota \& van de Bruck (2004), Horellou \& Berge (2005), 
Zeng \& Gao (2005), Maor \& Lahav (2005), Bartelmann, Doran 
\& Wetterich (2005), Nunes \& Mota (2006) and Wang (2006) in order to 
take into account models with a time
varying equation of state. Thus, in this section, we review only some basic 
concepts of the problem. Within the framework of the spherical
collapse model we assume a spherical mass overdensity shell, 
utilizing both the
virial theorem $T_{c}=-\frac{1}{2}U_{Gc}+U_{Qc}$ and the energy 
conservation $T_{c,f}+U_{Gc,f}+U_{Qc,f}=U_{Gc,t}+U_{Qc,t}$ where,
$T_{c}$ is the kinetic energy, $U_{Gc}=-3GM^{2}/5R$ is 
the potential energy and $U_{Qc}$
is the potential energy associated with the dark energy for the spherical
overdensity. In particular, the potential energy induced by the 
dark energy component (see Horellou \& Berge 2005 and references therein) is
\begin{equation}
U_{Qc}=(1+3w)\rho_{Qc}\frac{4\pi GM}{10}R^{2} \;\;\;.
\end{equation}
Using the above formulation  
we can obtain a cubic equation which relates the ratio 
between the final (virial) $R_{\rm f}$ and 
the turn-around outer radius $R_{\rm t}$:
\begin{equation}
2n_{1} \lambda^{3}-(2+n_{2})\lambda+1=0 
\end{equation}
where $\lambda=\frac{R_{\rm f}}{R_{\rm t}}$ is the 
collapse factor, 
\begin{equation}
n_{1}=-(3w+1)\frac{\Omega_{Q} f(\alpha_{\rm f})} 
{\zeta \Omega_{\rm m} \alpha_{\rm t}^{-3} }
\end{equation} 
and 
\begin{equation}
n_{2}=-(3w+1)\frac{\Omega_{Q}f(\alpha_{\rm t})} 
{\zeta \Omega_{\rm m} \alpha_{\rm t}^{-3}}
\end{equation} 
with
\begin{equation}
\zeta\equiv \frac{\rho_{\rm mc, t}}{\rho_{\rm m, t}}=
\left(\frac{R_{\rm t}}{\alpha_{\rm t}}\right)^{-3}
\end{equation}
(the definition of the $\alpha_{\rm t}$ and 
$\alpha_{\rm f}$ factors are presented bellow).  

However, we would like to point out that there is some confusion in 
the literature regarding eq.(9) which is based on energy
conservation. Indeed recently
it has been shown (Maor \& Lahav 2005) that the above formulation is
problematic due to the fact that the total energy of the bound
system is not conserved in dark energy models with $w\ne -1$. In order to 
avoid these systematic effects in this work we utilize 
two different assumptions. First of all, we assume that 
when the matter 
epoch just begins, the dark energy moves synchronously 
with ordinary matter on both, the Hubble scale and the galaxy cluster 
scale, the so called clustered dark energy scenario
(see Zeng \& Gao 2005 and references therein). 
It is interesting to mention that the synchronous dark energy 
evolution (hereafter clustered) was designed 
to conserve energy. 
In particular, Maor \& Lahav (2005) address the issue of the
clustered dark energy model based on the following assumptions: (i)
clustered quintessence considering that the whole system virializes
(matter and dark energy) and (ii) the dark energy remains clustered
but now only the matter virializes
(for more details see their section 4). Note, that in this work we    
are utilizing the second possibility.
In that case the equation which defines the collapse 
factor becomes:
\begin{equation}
(1+q)\lambda-\frac{q}{2}(1-3w)\lambda^{-3w}=\frac{1}{2}
\end{equation}
 where the factor $q$ is given by
\begin{equation}
q=\frac{\rho_{\rm Q,t}}{\rho_{\rm mc,t}}=\frac{\rho_{\rm Q,t}}{\zeta
  \rho_{\rm m,t}}=\frac{\nu}{\zeta}
\end{equation}
(the definition of the $\nu$ parameter is presented in section 3.1).  

If we assume that the dark energy component in a galaxy cluster scale 
can be treated as being homogeneous then we get the following 
cubic equation defined by Maor \& Lahav (2005) 
\begin{equation}
2q\lambda^{3}-(1+q)\lambda+\frac{1}{2}=0\;\;,
\end{equation}
Although it was pointed out recently (see Wang 2006) that if the value
of $q$ is less than 0.01 
then the problem of energy conservation
does not really affect the virialization process and thus 
eq.(9) is still a good approximation.  

From now on, we will call $\alpha_{\rm t}$ the scale factor of the
universe where the overdensity reaches at its maximum expansion and 
$\alpha_{\rm f}$ the scale factor in which the sphere virializes,
while $R_{\rm t}$ and $R_{\rm f}$ the corresponding radii of the spherical
overdensity. 
Note that
$\rho_{\rm mc, t}$ is the matter density which contains the 
spherical overdensity at the turn around time while 
$\rho_{\rm m, t}$ is the background matter density at the same epoch.
Therefore, for
bound perturbations which do not expand forever, the time needed 
to re-collapse 
is twice the turn-around time, $t_{\rm f}=2t_{\rm t}$. Doing so 
and taking into account eq.(4), it is easy to estimate the relation 
between the $\alpha_{\rm f}$ and $\alpha_{\rm t}$ 
\begin{equation}
\int_{0}^{\alpha_{\rm f}}  
\frac{1}{H(\alpha) \alpha} {\rm d}\alpha
=2  
\int_{0}^{\alpha_{\rm t}}  
\frac{1}{H(\alpha) \alpha} {\rm d}\alpha \;\;.
\end{equation}

In order to reduce the parameter space of the overall problem 
we put further observational constrains. 
Indeed, 
$\alpha_{\rm f}$ can be defined from the literature assuming that 
clusters have collapsed prior to the epoch of 
$z_{\rm f}\simeq 1.4$ in which 
the most distant cluster has been found (Mullis et al. 2005; Stanford
et al. 2006).
Therefore, considering $\alpha_{\rm f}\simeq 0.41$ and utilizing
eq.(16), it is routine to obtain the scale factor at the 
turn around time\footnote{The epoch of the 
turn around is roughly $z_{\rm t}\sim 2.5$}.   
In the case of a $\Lambda$ cosmology we get an analytical solution:
\begin{equation}
{\rm sin^{-1}h}\left(a_{\rm f}^{3/2} \sqrt{\nu_{0}} \right)=
2{\rm sin^{-1}h}\left(a_{\rm t}^{3/2} \sqrt{\nu_{0}} \right)
\end{equation} 
where $\nu_{0}=(1-\Omega_{\rm m})/\Omega_{\rm m}$.
While for the general problem we have to 
solve equation (16) numerically. The ratio between the 
scale factors converges to the Einstein de Sitter value
$\left(\frac{\alpha_{\rm f}}{\alpha_{\rm t}}\right)=2^{2/3}$ 
at high redshifts.

\subsection{The rescaled equations}
In this section we present some of the basic concepts of the spherical
collapse model. In particular, we assume a spherical 
overdensity which contains a
dark energy component which behaves either as clustered 
or homogeneous.
For a flat cosmology ($k=0$), 
using the basic differential equations (see eq.1 and 7) 
and performing the 
following transformations
\begin{equation}
x=\frac{\alpha}{\alpha_{\rm t}}\;\;\;\; {\rm and}\;\;\;\; 
y=\frac{R}{R_{\rm t}} \;\;\;,
\end{equation}
the evolution of the scale factors of the background and 
of the perturbation are 
governed respectively by the following two equations:

\begin{equation}
{\dot x}^{2}=H_{\rm t}^{2}\Omega_{\rm m,t}
\left[ \Omega_{\rm m}(x) x \right]^{-1}
\end{equation}
and 
\begin{equation}
{\ddot y}=-\frac{H_{\rm t}^{2}\Omega_{\rm m,t}}{2}
\left[ \frac{\zeta}{y^{2}}+\nu y I(x,y)\right] \;\;\;. 
\end{equation}

Note, that in order to obtain the above
set of equations we have used the following relations:
\begin{equation}
\rho_{\rm mc}=\rho_{\rm mc, t} \left(\frac{R}{R_{\rm t}}\right)^{-3}=
\frac{\zeta \rho_{\rm m, t}}{y^{3}}
\end{equation}
and
\begin{equation}
I(x,y)=\left\{ \begin{array}{cc}
       \left[1+3w(R(y))\right]\frac{f(R(y))}{f(\alpha_{\rm t})} &
       \mbox{Clustered DE}\\
       \left[1+3w(x)\right]f(x) & \mbox{Homogeneous DE}
       \end{array}
        \right.
\end{equation}
with
\begin{equation}
\nu=\frac{\rho_{\rm Q, t}}{\rho_{\rm m, t}}=\frac{1-\Omega_{\rm m,
    t}}{\Omega_{\rm m, t}} \;\;\;.
\end{equation}
The function $R(y)$ is given by 
the combination of eq.(12) and eq.(18): 
$R(y)=R_{\rm t}y=\zeta^{-1/3}\alpha_{\rm t}y$.
Finally, $\Omega_{\rm m}(x)$ is given by
\begin{equation}
\Omega_{\rm m}(x)=\frac{1}{1+\nu x^{3} f(x)} \;\;.
\end{equation}

\subsection{The general solution for the clustered dark energy scenario}
In this section we present our
analytical solution of the $\zeta$ parameter by integrating the
above system of differential equations, (eq.19 and eq.20), 
using at the same time the boundary conditions 
of $({\rm  d} y/{\rm d} x)_{x=1}=0$ and $y=1$. 
In the case of a homogeneous dark energy the above system is
solved only {\it numerically}. The novelty of our approach here is that 
for the clustered dark energy case
the system can be solved analytically including models, 
where the dark energy parameter $w$ is a function of the
cosmic time.
Indeed, due to the fact that now the second part of eq.(20) 
is a function only of $y$ we can perform easily the integration 

\begin{equation}
{\dot y}^{2}=-H_{\rm t}^{2}\Omega_{\rm m, t}\left[-\zeta y^{-1}+G(y)-C \right]
\end{equation}
where $C$ is the integration constant and
$$G(y)=
\frac{\nu}{f(\alpha_{\rm t})} \int y \left[1+3w(R(y))\right] f(R(y)){\rm
  d}y \;\; .$$
Considering now that the functional form of $f(R(y))$ is exponential 
(see the appendix) we have the following useful formula: 

\begin{equation}
\begin{array}{rcl} 
\int y [1+3w(R(y))] f(R(y)){\rm d}y =-2\int y f(R(y)){\rm d}y \\ 
-\int y^{2} \frac{{\rm d}f(R(y))}{{\rm d} y} {\rm d}y  
=-y^{2}f(R(y))\;\;\;.
\end{array}  
\end{equation} 
Therefore, the basic differential equation for the evolution 
of the overdensity perturbations under the 
framework of the boundary conditions [described before $C=\zeta+P(1)\nu$]
takes the form
\begin{equation}
\left(\frac{{\rm d}y}{{\rm d}x}\right)^{2}=
\frac{ \zeta y^{-1}+\nu P(y)-\zeta-P(1)\nu}
{\left[x\Omega_{\rm m}(x)\right]^{-1}}\;\;,
\end{equation}
with 
\begin{equation}
P(y)=\frac{y^{2}f(R(y))}{f(\alpha_{\rm t})} \;\;.
\end{equation}

Finally, below we present the general integral equation which
governs the behavior of the density contrast $\zeta$ at the turn
around epoch, for generic flat
cosmological models, as a function of the equation of state (which
also depends on time) and the perturbation collapse time
\begin{equation}
\begin{array}{rcl} 
\int_{0}^{1} \left[\frac{y}{\zeta+\nu yP(y)-(\zeta+P(1)\nu)y}\right]^{1/2}
    {\rm d}y \\ 
=\int_{0}^{1} \left[x\Omega_{\rm m}(x)\right]^{1/2}
{\rm d}x
\end{array}
\end{equation}

\section{Specific Dark Energy models}
Knowing the functional form of the function $f(x)$, the collapse scale
factor (in our case $\alpha_{\rm f}\simeq 0.41$), the turn around scale 
factor $\alpha_{\rm t}$ (from eq. 16) and assuming a 
low matter density flat cosmological 
model with $\Omega_{\rm m}=1-\Omega_{\rm Q}=0.3$, we can obtain
the parameter $\zeta$ solving numerically the equation (29).
Note that in this work, we deal with four different 
type of equations of state, which we present in the coming sections.

\begin{figure}
\mbox{\epsfxsize=12cm \epsffile{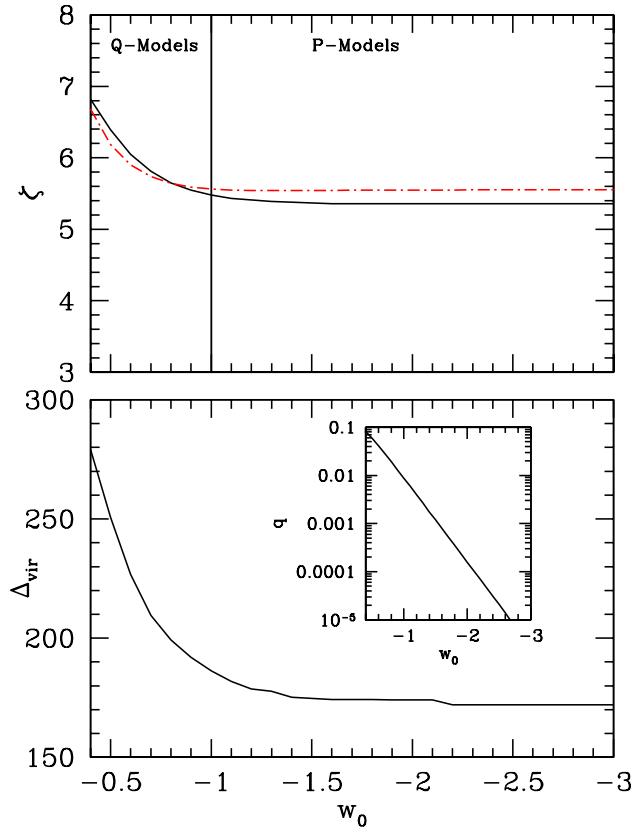}}
\caption{The upper panel shows the $\zeta$ parameter as a function of
  $w_{0}$. Note that the solid line presents
the clustered dark energy scenario while the dashed line corresponds
to the homogeneous dark energy. The 
lower panel shows the corresponding 
density contrast at virialization, for a flat cosmology, $\Omega_{\rm
  m}=1-\Omega_{Q}=0.3$ with a constant equation of state. Finally, the
insert plot shows the behavior of the $q$ parameter for different 
$w_{0}$. Finally, the index P and Q reads Phantom and Quintessence
respectively.}
\end{figure}

\subsection{Quintessence - Phantom models}
In this case the equation of state is constant (see for a review, 
Peebles \& Ratra 2003; Caldwell 2002; Corasaniti et al. 2004). 
If $-1\le w_{0}<-1/3$ we have the so called 
Quintessence models while for $w_{0}<-1$ we get the so called Phantom 
cosmologies (hereafter QP-models). Despite the fact 
that for these models the cluster virialization has been 
investigated thoroughly in several papers we have decided to re-estimate 
it in order to understand the variants from the $w=w(t)$ case.
In particular, for
$w(\alpha)=w_{0}$ we get $f(\alpha)=\alpha^{-3(1+w_{0})}$ 
and the overall problem (eq. 29) reduces to the
integral equation found by Zeng \& Gao (2005). In particular, 
the basic factors become
\begin{equation}
P(y)=\zeta^{1+w_{0}}y^{-3w_{0}-1}
\end{equation}
and 
\begin{equation}
\Omega_{\rm m}(x)=\frac{1}{1+\nu x^{-3w_{0}}} \;\;.
\end{equation}
In Figure 1 we present for the clustered (solid line) and 
homogeneous dark energy respectively (dashed line) 
the $\zeta$ solution (top panel),
as a function of the equation 
of state parameter and it is obvious that the two models give
almost the same results\footnote{For $w_{0}<-0.9$ the
$\zeta$ solution tends to the Einstein-de Sitter case 
$\zeta \longrightarrow \left(\frac{3 \pi}{4}\right)^{2}$}. Also 
Figure 1 (bottom panel) shows the dependence of the 
density contrast at virialization on $w$.
 
\begin{equation}
\Delta_{\rm vir}=\frac{\rho_{\rm mc, f}}{\rho_{\rm m, f}}=
\frac{\zeta}{\lambda^{3}} 
\left(\frac{\alpha_{\rm f}}{\alpha_{\rm t}}\right)^{3}
\end{equation}
where $\rho_{\rm mc, f}$ is the matter density in the virialized structure 
while $\rho_{\rm m, f}$ is the background matter density 
at the same epoch.
The factor $\Delta_{\rm vir}$ is a key parameter in this kind of
studies because we can compare the predictions 
provided by the spherical collapse model with observations.
In this framework, it is also interesting to note that 
the collapse factor 
is in between $0.48\le \lambda \le 0.50$ while for the homogeneous dark
energy we get $0.47\le \lambda \le 0.50$. Both cases tend to the 
standard value $0.5$ in an Einstein de Sitter universe
in agreement with previous studies 
(Maor \& Lahav 2005; Wang 2006). This is
to be expected simply because at large redshifts matter dominates 
over the dark energy in the universe which means that 
the parameter $q$ has small values (see the insert plot of Fig. 1).
Note that $\Lambda$-models can be described by 
quintessence models
with $w_{0}$ strictly equal to -1 and thus, eq.(29) is written:

\begin{equation}
\int_{0}^{1} \left[\frac{y}{\zeta+\nu y^{3}-(\zeta+\nu)y}\right]^{1/2}
    {\rm d}y=\frac{{\rm ln}(\sqrt{1+\nu}+\sqrt{\nu})^{2/3}}{\sqrt{\nu}}\;\;.
\end{equation}

\begin{figure}
\mbox{\epsfxsize=12cm \epsffile{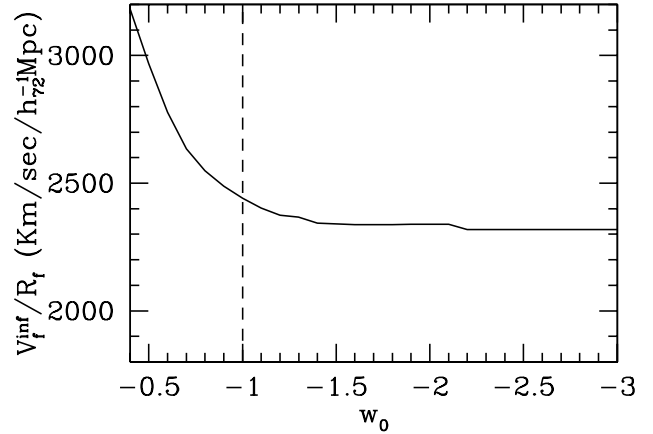}}
\caption{The non linear infall velocity in the virialized regime as 
a function of the equation of state parameter.}
\end{figure}

In order to explore further the virialized structures we 
investigate the connection between the infall velocity at the virialized 
epoch and the equation of state. It is well known from the literature 
that the non linear infall velocity field is described by the following 
expression (see Lilje \& Efstathiou 1989; Croft, 
Dalton \& Efstathiou 1999): 
\begin{equation}
V^{\rm inf}(r)=-\frac{1}{3}H(\alpha) \Omega_{\rm m}^{\beta}(\alpha)r 
\frac{\delta(r)}{[1+\delta(r)]^{0.25}} \;\;, 
\end{equation} 
where $\delta(r)$ is the fluctuation field. It should be noted 
that in this work we have used 
the generic expression for 
the growth index ($\simeq \Omega_{\rm m}^{\beta}$) 
defined by Wang \& Steinhardt (1998)
with 
\begin{equation}
{\beta} \simeq \frac{3}{5-w_{0}/(1-w_{0})}+
\frac{3(1-w_{0})(1-3w_{0}/2)}
{125(1-6w_{0}/5)^{3}}(1-\Omega_{\rm m}) \;.  
\end{equation}
Due to interplay between the infall velocity and the 
virial radius, in Figure 2 we present the corresponding 
ratio between them at 
the virialized epoch, [$r=R_{\rm f}$, $\alpha=\alpha_{\rm f}$ and
$\delta(R_{\rm f})\simeq \Delta_{\rm vir}$], 
as a function of the equation of state parameter. 
Therefore, it becomes evident that for models with $w_{0}<-1$,
the so called Phantom models (see Caldwell 2002), the 
infall velocity is unaffected really by the equation of state. When
we use models with $w>-1$ the infall velocity is a decreasing function 
of $w$ (the difference is $\sim 37\%$). The latter 
is to be expected because of the $\Delta_{\rm vir}$ 
behavior (see Figure 1).

\subsection{Models with time variation $w$}
The last few years there have been many theoretical 
speculations regarding the nature 
of the exotic ``dark energy''. Various candidates 
have been proposed in the literature, among which 
a dynamical scalar field acting as vacuum energy
(Ozer \& Taha 1987). Under this framework, high energy field 
theories generically show that the equation of state of such
a dark energy is a function of the cosmic time.
If this is the case then the basic equations regarding the 
cluster virilization in models with a time varying equation of state
become more complicated than in models with constant $w$.
Note that in the literature due to the absence of a physically
well-motivated fundamental theory, there is plenty of dark 
energy models that can 
fit current observational data (for a review see 
Liberato \& Rosenfeld 2006). In this work as an example, 
we use several parameterizations regarding the dark
energy component  
in order to evaluate the integrals of eq.(3) and eq. (29). In particular, we
consider a simple expression for the equation of state:

$$w(\alpha)=w_{0}+w_{1}X(\alpha)$$ 
which is available in the literature
(see for example Goliath et al. 2001; 
Linder 2003; Cepa 2004; Liberato \& Rosenfeld 2006 
and references therein). The function $X(\alpha)$ satisfies 
the following constrain at the present epoch: $X(1)=0$
and thus, $w_{0}<-1/3$. Here we sample the parameters as follows:
$w_{0} \in [-1.05,-0.35]$ and $w_{1} \in [0.1,1.5]$ in steps of 0.1. 
To this end, we need
to emphasize that 
our approach is powerful in a sense that for different parametric
forms of $w(\alpha)$ the system of eq. (19) and
eq. (20) can be solved analytically in the framework of a clustered
dark energy. Finally, in this work we assume that 
the parameters $(w_{0},w_{1})$ and the functional form of the equation
of state parameter remain the same either for the
entire universe or for the spherical perturbations.

\subsubsection{The $w(\alpha)=w_{0}+w_{1}(1-\alpha)$ model}
In this case, (see Linder 2003; Cepa 2004) where $w_{0}$ and
$w_{1}$ are constants, using eq.(3), eq.(24)
and eq.(28) we derive the following basic functions of the 
differential equation eq.(29):

\begin{equation}
f(\alpha)=\alpha^{-3(1+w_{0}+w_{1})}e^{3w_{1}(\alpha-1)}
\end{equation}

\begin{equation}
\Omega_{\rm m}(x)=
\frac{1} { 1+\nu x^{-3(w_{0}+w_{1})}e^{3w_{1}(x-1)} }
\end{equation}
and 
\begin{equation}
P(y)=\frac{\zeta^{1+w_{0}+w_{1}} e^{3w_{1}\alpha_{\rm
      t}(\zeta^{-1/3}y-1)} }{y^{3(w_{0}+w_{1})+1}}\;\;.
\end{equation}

\begin{figure}
\mbox{\epsfxsize=12cm \epsffile{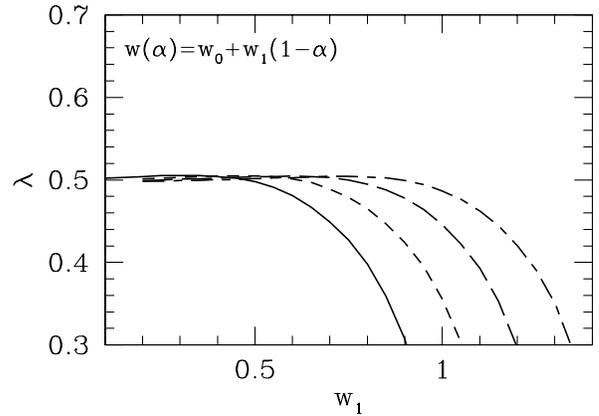}}
\caption{Possible pairs $(w_{0},w_{1}$) defining the 
collapse factor $\lambda$ for the clustered case, 
based on $w(\alpha)=w_{0}+w_{1}(1-\alpha)$ [see section 4.2.1].
The lines correspond to $w_{0}=-0.35$ (solid), $w_{0}=-0.45$ (short dashed),
$w_{0}=-0.55$ (dashed) and $w_{0}=-0.65$ (dot dashed).}
\end{figure}

For the homogeneous dark energy scenario we find that the collapse factor
lies in the range $0.47\le \lambda \le 0.50$ while for the clustered
case we get $0.30\le \lambda \le 0.50$. In particular, 
Fig.3 shows the behavior of the collapse factor for 
possible pairs of $(w_{0},w_{1}$) starting from $w_{0}=-0.35$ (solid
line), $w_{0}=-0.45$ (short dashed line), $w_{0}=-0.55$ (dashed line) and
$w_{0}=-0.65$ (dot dashed line). 
From the figure, it becomes 
evident that for large values of $w_{1}$ the 
collapse factor is significant less than 0.5 which means that
under the framework of the present equation of state,
a candidate structure (large scale overdensity) is 
located in a large density regime ($\Delta_{\rm
  vir}>400$) and at the end the tendency is to
collapse in a more bound system, with respect to the
QP-models ($w=const$). 
Additionally, it is interesting to
mention that for small values of $w_{1}<0.55$ the structures 
reveal a similar dynamical behavior, $0.48\le \lambda \le 0.50$,
as described in section 4.1.

\subsubsection{The $w(\alpha)=w_{0}+w_{1}(\alpha^{-1}-1)$ model}
Using now the equation of state derived by 
Goliath et al. (2001) then the formulas
described before become: 
\begin{equation}
f(\alpha)=\alpha^{-3(1+w_{0}-w_{1})}e^{3w_{1}(\alpha^{-1}-1)}
\end{equation}

\begin{equation}
\Omega_{\rm m}(x)=
\frac{1} { 1+\nu x^{-3(w_{0}-w_{1})}
e^{3w_{1}(x^{-1}-1)} }
\end{equation}
and 
\begin{equation}
P(y)=\frac{\zeta^{1+w_{0}-w_{1}} e^{3w_{1}\alpha^{-1}_{\rm
      t}(\zeta^{1/3}y^{-1}-1)} }{y^{3(w_{0}-w_{1})+1}}\;\;.
\end{equation}
Following the same paradigm as in the previous case 
in Fig.4 we present for the clustered case the 
distribution of $\lambda$. 
It is obvious that also this equation of state 
affects the virialization of the large scale structures 
in the same way as before (see section 4.2.1), 
but the corresponding $w_{1}$ parameter takes much lower values. 
For $w_{1}<0.15$ the collapse factor tends to 0.5. Note, that
for the homogeneous dark energy the collapse factor is 
in the range $0.45 \le \lambda \le 0.49$.  

\begin{figure}
\mbox{\epsfxsize=12cm \epsffile{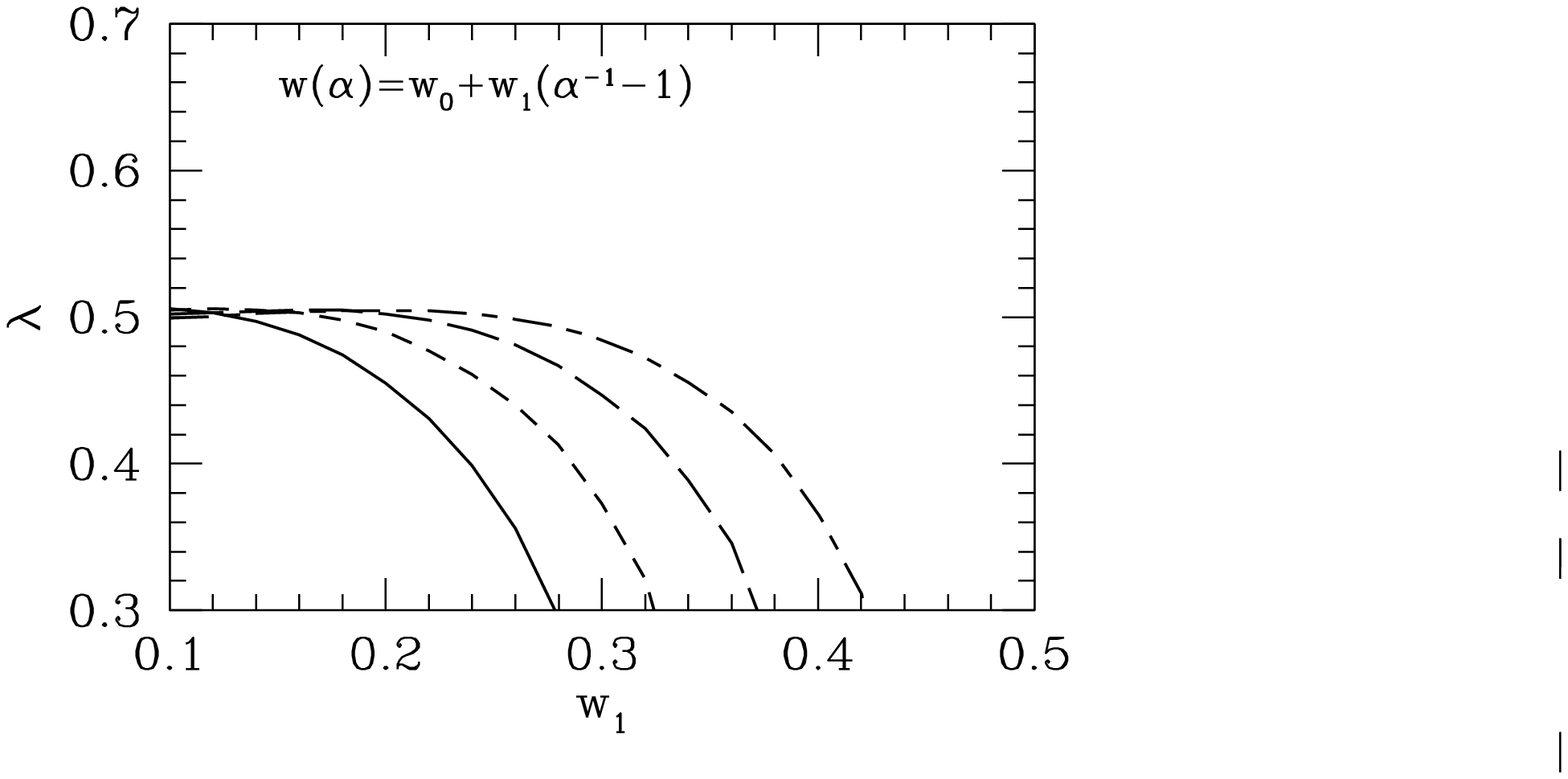}}
\caption{The collapse factor $\lambda$, 
based on $w(\alpha)=w_{0}+w_{1}(\alpha^{-1}-1)$ [see section 4.2.2].
The lines correspond to the same values of $w_{0}$ as in figure 3.}
\end{figure}

\subsubsection{The $w(\alpha)=w_{0}+w_{1}(1-\alpha)\alpha$ model}
In this case, (see Liberato \& Rosenfeld 2006) 
we get: 
\begin{equation}
f(\alpha)=\alpha^{-3(1+w_{0})}e^{\frac{3w_{1}}{2}
(\alpha-1)^{2}}
\end{equation}
\begin{equation}
\Omega_{\rm m}(x)=
\frac{1} { 1+\nu x^{-3w_{0}}
e^{\frac{3w_{1}}{2}(x-1)^{2}} }
\end{equation}
and 
\begin{equation}
P(y)=\frac{\zeta^{1+w_{0}} e^{\frac{3w_{1}}{2}\alpha_{\rm
      t}(\alpha_{\rm t}\zeta^{-2/3}y^{2}-\alpha_{\rm
      t}-2\zeta^{-1/3}y+2)} 
}{y^{3w_{0}+1}}\;\;.
\end{equation}
Performing once again the same constrains as before 
the virialization 
radius to the turn around radius 
lies in the range $0.46\le \lambda \le 0.50$. This result is 
in agreement with
those derived in the QP-models, 
despite the fact that the two dark energy models have completely 
different equations of state. Finally, considering now 
the homogeneous dark energy scenario the collapse factor is
almost in the same interval i.e, $0.45 \le \lambda \le 0.48$.

\section{Conclusions}
The launch of the recent observational data has brought great progress 
in understanding the physics of gravitational collapse as well as the 
mechanisms that have
given rise to the observed large-scale structure of the universe.
In this work assuming that the dark energy moves synchronously with
ordinary matter on both the Hubble scale and the galaxy cluster scale
(clustered dark energy), we have treated
analytically the nonlinear spherical collapse scenario considering
different models with negative pressure which contains also 
a time varying equation of state. We verify that having 
an overdense region of matter,
which will create a cluster of galaxies prior to the epoch 
of $z_{\rm f}\simeq 1.4$, dark energy affects the 
virialization of the large scale structures in the following manner:
flat cosmological models that contain either (i) a constant equation
state or (ii) an equation of state with
$w=w_{0}+w_{1}(1-\alpha)\alpha$ the
virialization radius divided by the turn around radius 
is in between $0.47\le R_{\rm f}/R_{\rm t} \le 0.50$ 
and the density contrast at virialization 
lies in the interval $171\le \Delta_{\rm vir} \le 240$.
It is interesting to mention that the same behavior is also found 
for the homogeneous dark energy.
Finally, we find that the following 
equations of state: $w=w_{0}+w_{1}(1-\alpha)$ 
and $w=w_{0}+w_{1}(\alpha^{-1}-1)$ correspond to  
relative larger $\Delta_{\rm vir}$'s and thus, 
produce more bound systems ($0.30\le R_{\rm f}/R_{\rm t} \le 0.50$) 
with respect to the other two equations of state 
mentioned before. While for the homogeneous dark energy we found
structures where the collapse factor is in between 
$0.45\le R_{\rm f}/R_{\rm t} \le 0.49$. 

\section*{Acknowledgments}
We would like to thank Manolis Plionis, Rien van de Weygaert, 
Bernard Jones and the anonymous referee for their useful comments and
suggestions. Spyros Basilakos has benefited from discussions 
with Joseph Silk. Finally, SB acknowledges support by 
the Nederlandse Onderzoekschool voor Astronomie
(NOVA) grant No 366243.

\section*{Appendix A}
Without wanting to appear too pedagogical, we remind the reader of some 
basic elements of Algebra. Given a cubic equation: 
$\lambda^{3}+a_{1} \lambda^{2}+a_{2} \lambda+a_{3}=0$,
let ${\cal } D$ be the discriminant:
\begin{equation}
{\cal D}=a_{1}^{2} a_{2}^{2}-4a_{2}^{3}-
4a_{1}^{3}a_{3}-27a_{3}^{2}+18a_{1}a_{2}a_{3}
\end{equation}  
and 
$$x_{1}=-a_{1}^{3}+\frac{9}{2}a_{1}a_{2}-\frac{27}{2}a_{3} \;,\;\;\;\;\;\;
x_{2}=-\frac{3\sqrt{3 \cal{D}}}{2}\;\;.$$
If ${\cal D}>0$, all roots are real (irreducible case). In that 
case $\lambda_{1}$, $\lambda_{2}$ and $\lambda_{3}$ can be written:

\begin{equation}
\lambda_{\mu}=-\frac{a_{1}}{3}-
\frac{2r^{1/3}}{3} {\rm cos} \left[\frac{\theta-(\mu-1)\pi}{3} \right]
\;\;\;\;\;\; \mu=1,2,3
\end{equation} 
where $r=\sqrt{x_{1}^{2}+x_{2}^{2}}$ and $\theta={\rm cos^{-1}} (x_{1}/r)$. 

In this study, we derive analytically the 
exact solution of the basic cubic equation (eq.15), 
having polynomial 
parameters: $a_{1}=0$, $a_{2}=-(1+q)/2q$ and
$a_{3}=1/4q$. 
Then the discriminant of eq.(15) is:
$${\cal D}(q)=\frac{8(1+q)^{3}-27q^{2}}{16q^{3}} \;\;.$$
Of course, in order to obtain physically acceptance results 
we need to take $q>0$ which gives ${\cal D}(q)>0$. 
Therefore, all roots of the cubic equation are real 
(irreducible case) but one of them $\lambda_{3}$
corresponds to expanding shells. 
It is obvious that for $q\longrightarrow 0$ the above
solution tends to the Einstein-de Sitter case 
($\lambda_{3}\longrightarrow 0.50$), as it should.

\section* {Appendix B}
We present here some more details regarding the integration of
eq. (26). Considering a spherical overdensity with radius $R$ 
the functional form of eq.(3) becomes:

\begin{equation}
f(R)={\rm exp}\left[ 3\int_{R}^{1} \left(
  \frac{1+w(u)}{u}\right) {\rm d}u \right] \;\;.
\end{equation}
Taking now into account that
$R(y)=R_{\rm t}y=\zeta^{-1/3}\alpha_{\rm t}y$, 
(described well in section 3.1) 
and differentiating the above integral we have the following useful 
formula: 
\begin{equation}
\frac{{\rm d}f(R(y))} {{\rm d} y}=\frac{{\rm d}f}{{\rm d}R}
\frac{{\rm d}R}{{\rm d}y}=
-3y^{-1}[1+w(R(y))]f(R(y)) \;.
\end{equation} 
Using now the integral of eq. (26)
 \begin{equation}
\begin{array}{rcl} 
\int y [1+3w(R(y))] f(R(y)){\rm d}y=-2\int y f(R(y)){\rm d}y\\
+3\int[1+w(R(y))]yf(R(y)){\rm d}y\;\;
\end{array} 
\end{equation}   
and taking into account eq. (48) we can easily solve the integral
 \begin{equation}
\begin{array}{rcl} 
\int y [1+3w(R(y))] f(R(y)){\rm d}y=-2\int y f(R(y)){\rm d}y\\
-\int y^{2} \frac{{\rm d}f(R(y))}{{\rm d} y} {\rm d}y=-y^{2}f(R(y))\;\;
\end{array} 
\end{equation}


{\small 
\begin{thebibliography}{}
\bibitem[]{}Bartelmann M., Doran M., \&, Wetterich, C., 2006, 
A\&A, 454, 27
\bibitem[]{}Basilakos S., 2003, ApJ, 590, 636
\bibitem[]{}Basilakos S., \&, M. Plionis M., MNRAS, 2005 
360, L35
\bibitem[]{}Battye R. A., \&, Weller J., Phys. Rev. D, 2003, 68, 3506
\bibitem[]{}Caldwell R. R., Dave R., Steinhardt P. J.,  
Phys. Rev. Lett., 1998, 80, 1582
\bibitem[]{}Caldwell, R. R., Phys. Let. B, 2002, 545, 23
\bibitem[]{}Cepa J., A\&A, 2004, 422, 831
\bibitem[]{}Corasaniti, P. S., Kunz, M., Parkinson, D.,
  Copeland. E. J., \&, Bassett, B. A., Phys. Rev. D, 2004, 70, 3006
\bibitem[]{}Croft, R. A. C., Dalton, G. B., \&, 
Efstathiou, G., MNRAS, 1999, 305, 547
\bibitem[]{}Freedman W. L., et al., ApJ, 2001, 553, 47 
\bibitem[]{}Gunn J. E., \&, Gott J. R., ApJ, 1972 
176, 1
\bibitem[]{}Goliath A., Amanulah R., Astier P., Goobar A, \&, 
Pain R., A\&A, 2001, 380, 6
\bibitem[]{}Horellou C., \&, Berge J., 2005, MNRAS, 360, 1393
\bibitem[]{}Iliev I. T., \&, Shapiro P. R., 2001, MNRAS, 325, 468
\bibitem[]{}Lahav O., Lilje P. B., Primack J. R., \&, Rees M. J., 
MNRAS, 1991, 251, 128
\bibitem[]{}Liberato, L., \&, Rosenfeld, R., 2006, JCAP, 7, 9 
\bibitem[]{}Lilje P. B., \&, Efstathiou, G., MNRAS, 1989, 236, 851 
\bibitem[]{}Linder E. V., Phys. Rev. Lett., 2003, 90, 1301
\bibitem[]{}Lokas E. L., Acta Physica Polonica B, 2001, 32, 3643
\bibitem[]{}Mainini, R., Maccio, A. V., Bonometto, S. A., \&, Klypin,
  A., ApJ, 2003, 599, 24, 
\bibitem[]{}Maor I., \&, Lahav O., Journal of Cosmology and
  Astroparticle Physics, 2005, 7, 3 
\bibitem[]{}Mota D. F., \&, van de Bruck C., A\&A, 2004, 421, 71
\bibitem[]{}Mullis C. R., Rosati P., Lamer G., B$\ddot {\rm o}$ehringer H.,
Schuecker P., \&, Fassbender R., MNRAS, 2005, 623, L85
\bibitem[]{}Nunes N. J., \&, Mota D. F., MNRAS, 2006, 368, 751
\bibitem[]{}Ozer, M., \&, Taha, O., Nucl. Phys. B, 1987, 287,
  776
\bibitem[]{}Peebles P. J. E., \&, Ratra B., RvMP, 2003, 75, 559
\bibitem[]{}Percival W. J., A\&A, 2005, 443, 819
\bibitem[]{}Ratra B., \&, Peebles P. J. E., 
Phys. Rev. D, 1988, 37, 3406
\bibitem[]{}Spergel D. N., et al., ApJS, 2003, 148, 175
\bibitem[]{}Spergel D. N., et al., ApJ, 2006, submitted,
  (astro-ph/0603449)
\bibitem[]{}Stanford, S. A., et al., ApJ, 2006, 646, L13
\bibitem[]{}Tegmark M., et al. , Phys. Rev. D, 2004, 69, 3501
\bibitem[]{}Wang L., \&, Steinhardt P. J., ApJ, 1998, 
508, 483
\bibitem[]{}Wang P., ApJ, 2006, 640, 18
\bibitem[]{}Weinberg N. N., \&, Kamionkowski M., ApJ, 2003, 341,
  251
\bibitem[]{}Zeng, Ding-fang, \&, Gao, Yi-hong., 2005, (astro-ph/0505164)

\end{thebibliography}
}
\end{document}